\documentclass[twoside,twocolumn,9pt]{article}
\usepackage{extsizes}
\usepackage[super,sort&compress,comma]{natbib} 
\usepackage[version=3]{mhchem}
\usepackage[left=1.5cm, right=1.5cm, top=1.785cm, bottom=2.0cm]{geometry}
\usepackage{balance}
\usepackage{times,mathptmx}
\usepackage{sectsty}
\usepackage{graphicx} 
\usepackage{lastpage}
\usepackage[format=plain,justification=justified,singlelinecheck=false,font={stretch=1.125,small,sf},labelfont=bf,labelsep=space]{caption}
\usepackage{float}
\usepackage{fancyhdr}
\usepackage{fnpos}
\usepackage[english]{babel}
\addto{\captionsenglish}{%
  
}
\usepackage{array}
\usepackage{droidsans}
\usepackage{charter}
\usepackage[T1]{fontenc}
\usepackage[usenames,dvipsnames]{xcolor}
\usepackage{setspace}
\usepackage[compact]{titlesec}
\usepackage{hyperref}

\usepackage{epstopdf}

\definecolor{cream}{RGB}{222,217,201}

\begin{document}

\pagestyle{fancy}
\thispagestyle{plain}
\fancypagestyle{plain}{

\renewcommand{\headrulewidth}{0pt}
}

\makeFNbottom
\makeatletter
\renewcommand\LARGE{\@setfontsize\LARGE{15pt}{17}}
\renewcommand\Large{\@setfontsize\Large{12pt}{14}}
\renewcommand\large{\@setfontsize\large{10pt}{12}}
\renewcommand\footnotesize{\@setfontsize\footnotesize{7pt}{10}}
\makeatother

\renewcommand{\thefootnote}{\fnsymbol{footnote}}
\renewcommand\footnoterule{\vspace*{1pt}%
\color{cream}\hrule width 3.5in height 0.4pt \color{black}\vspace*{5pt}} 
\setcounter{secnumdepth}{5}

\makeatletter 
\renewcommand\@biblabel[1]{#1}            
\renewcommand\@makefntext[1]%
{\noindent\makebox[0pt][r]{\@thefnmark\,}#1}
\makeatother 
\renewcommand{\figurename}{\small{Fig.}~}
\sectionfont{\sffamily\Large}
\subsectionfont{\normalsize}
\subsubsectionfont{\bf}
\setstretch{1.125} 
\setlength{\skip\footins}{0.8cm}
\setlength{\footnotesep}{0.25cm}
\setlength{\jot}{10pt}
\titlespacing*{\section}{0pt}{4pt}{4pt}
\titlespacing*{\subsection}{0pt}{15pt}{1pt}

\fancyfoot{}
\fancyfoot[RO]{\footnotesize{\sffamily{1--\pageref{LastPage} ~\textbar  \hspace{2pt}\thepage}}}
\fancyfoot[LE]{\footnotesize{\sffamily{\thepage~\textbar\hspace{3.45cm} 1--\pageref{LastPage}}}}
\fancyhead{}
\renewcommand{\headrulewidth}{0pt} 
\renewcommand{\footrulewidth}{0pt}
\setlength{\arrayrulewidth}{1pt}
\setlength{\columnsep}{6.5mm}
\setlength\bibsep{1pt}

\makeatletter 
\newlength{\figrulesep} 
\setlength{\figrulesep}{0.5\textfloatsep} 

\newcommand{\topfigrule}{\vspace*{-1pt}%
\noindent{\color{cream}\rule[-\figrulesep]{\columnwidth}{1.5pt}} }

\newcommand{\botfigrule}{\vspace*{-2pt}%
\noindent{\color{cream}\rule[\figrulesep]{\columnwidth}{1.5pt}} }

\newcommand{\dblfigrule}{\vspace*{-1pt}%
\noindent{\color{cream}\rule[-\figrulesep]{\textwidth}{1.5pt}} }

\makeatother

\twocolumn[
  \begin{@twocolumnfalse}
\vspace{3cm}
\sffamily
\begin{tabular}{m{4.5cm} p{13.5cm} }

\includegraphics{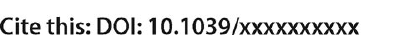} & \noindent\LARGE{\textbf{Confinement and activity regulate bacterial motion$~~~~$ in porous media}} \\
\vspace{0.3cm} & \vspace{0.3cm} \\

 & \noindent\Large{Tapomoy Bhattacharjee\textit{$^{a}$} and Sujit S. Datta\textit{$^{b}$$^{\ast}$}} \\
 
\includegraphics{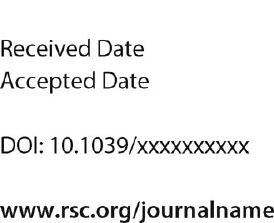} & \noindent\normalsize{Understanding how bacteria move in porous media is critical to applications in healthcare, agriculture, environmental remediation, and chemical sensing. Recent work has demonstrated that \textit{E. coli}, which moves by run-and-tumble dynamics in a homogeneous medium, exhibits a new form of motility when confined in a disordered porous medium: hopping-and-trapping motility, in which cells perform rapid, directed hops punctuated by intervals of slow, undirected trapping. Here, we use direct visualization to shed light on how these processes depend on pore-scale confinement and cellular activity. We find that hopping is determined by pore-scale confinement, and is independent of cellular activity; by contrast, trapping is determined by the competition between pore-scale confinement and cellular activity, as predicted by an entropic trapping model. These results thus help to elucidate the factors that regulate bacterial motion in porous media, and could help aid the development of new models of motility in heterogeneous environments. } \\

\end{tabular}

 \end{@twocolumnfalse} \vspace{0.6cm}

  ]

\renewcommand*\rmdefault{bch}\normalfont\upshape
\rmfamily
\section*{}
\vspace{-1cm}


\footnotetext{\textit{$^{a}$~The Andlinger Center for Energy and the Environment, Princeton University, 86 Olden Street, Princeton, NJ, 08544, USA\\
$^{b}$~Department of Chemical and Biological Engineering, Princeton University, Princeton, New Jersey, 08544, United States;\\ *E-mail: ssdatta@princeton.edu}}


\section{Introduction}
While studies of bacterial motility date back to as early as 1676, they typically focus on cells in homogeneous environments, such as in liquid culture and near flat surfaces. However, in most real-world settings, bacteria must navigate heterogeneous three-dimensional (3D) porous media such as gels, tissues, soils, and sediments. This process can be harmful: for example, during an infection, pathogens squeeze through pores in tissues and gels and thereby spread through the body.\cite{balzan,chaban,pnas,harman,ribet,siitonen,lux,oneil} Similarly, during meat spoilage, pathogenic bacteria squeeze through pores in tissue and spread in contaminated meat.\cite{Gill1977,Shirai2017} This process can also be beneficial: for example, a promising route toward cancer treatment relies on engineered bacteria penetrating into tumors and delivering anticancer agents.\cite{thornlow,toley} In agriculture, rhizosphere bacteria move through soils to help sustain and protect plant roots, which impacts crop growth and productivity.\cite{dechesne,souza,turnbull,watt,babalola} Further, in environmental settings, bioremediation efforts often seek to apply motile bacteria to migrate towards and degrade contaminants trapped in groundwater aquifers.\cite{roseanne18,roseanne15,ford07,wang08} However, despite their potentially harmful or beneficial consequences, there is an incomplete understanding of how bacteria---and more generally, how active particles---move in 3D porous media. This gap in knowledge hinders attempts to predict the spread of infections, design new bacterial therapies, and develop effective agricultural and bioremediation strategies. Moreover, while the diffusion of "passive", thermally-equilibrated particles in random media is well-studied, how self-propulsion or "activity" impacts migration through a porous medium is unknown---hindering our ability to control active particles in applications ranging from drug delivery to chemical sensing.\cite{Nelson2010,Gao2014a,Patra2013,Abdelmohsen2014,Wang2012a,Ebbens2016} 

In recent work,\cite{tapa19} we discovered that \textit{E. coli}, a canonical example of active particles that moves by run-and-tumble dynamics in homogeneous environments,\cite{berg} exhibits a new form of motility in porous media in which cells are intermittently and transiently trapped as they navigate the pore space. When a cell is trapped, it constantly reorients its body until it is able to escape; thus, we expect that the trapping durations depend on both the geometry of the porous medium and cellular activity. The cell then moves in a directed path through the pore space, a process we refer to as hopping, until it again becomes trapped in a tight or tortuous spot; thus, we expect that the hopping lengths are determined solely by the geometry of the medium. However, whether these hypotheses are correct, and how exactly hopping and trapping depend on pore-scale confinement and cellular activity, is still unknown. As a result, our ability to accurately model and predict how bacteria move in porous media is limited.

\begin{figure*}[h]
\centering
  \includegraphics[height=5.5cm]{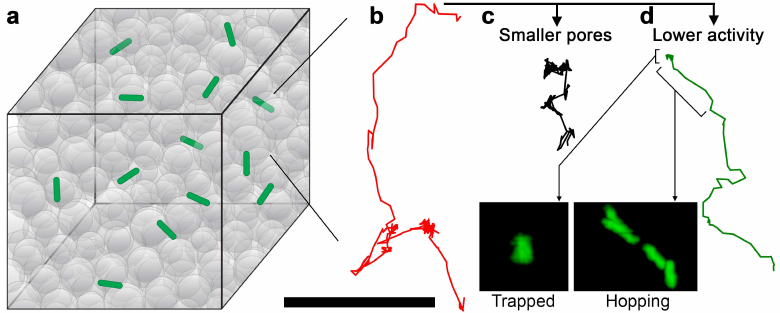}
  \caption{\textbf{Independently testing the influence of pore-scale confinement and cellular activity on bacterial motion in porous media. a} Schematic representation of a transparent, three-dimensional porous medium comprised of a jammed packing of hydrogel particles swollen in liquid cell culture medium (gray circles). Fluorescent \textit{E. coli} (green rods) swim through the pores between particles. \textbf{b} Representative trajectory of a cell moving through the pore space shows two distinct motility modes that the cell intermittently switches between: hopping, in which the cell moves rapidly along directed and extended paths in the pore space, and trapping, in which the cell is localized in tight spots of the pore space. \textbf{c} Representative trajectory of a cell in a medium with smaller pores, indicating that hopping lengths are smaller and trapping durations are longer. \textbf{d} Representative trajectory of a cell with lower activity, indicating that hopping lengths are unchanged but trapping durations are longer. Inset images show selected superimposed frames of cellular motion, equally spaced in time, showing that the cell is localized and randomly oriented during trapping, but moving and directed during hopping. All trajectories are for a time duration of $\approx26$ s. Scale bar is 20 $\mu$m. }
  \label{fgr:example}
\end{figure*}

In this Article, we study how pore-scale confinement and cellular activity individually influence bacterial trapping and hopping in 3D porous media. Consistent with our expectation, we find that the hopping lengths are determined by the geometry of the pore space, and are independent of cellular activity. By contrast, the trapping durations depend on both pore space geometry and cellular activity. Intriguingly, the distribution of trapping durations shows power-law scaling similar to that found for passive species in other disordered systems. The scaling exponent depends on the interplay between pore-scale confinement, which promotes trapping, and cellular activity, which enables cells to escape traps. These results thus expand on the tantalizing similarity between the motion of bacteria---which consume energy, and are thus out of thermal equilibrium---and a passive species navigating a disordered landscape. Ultimately, by shedding light on the physics underlying hopping and trapping, our work helps to provide a foundation by which bacterial motion in porous media can be predicted and even controlled.

\section{Results}
\subsection{Characterizing bacterial motion in 3D porous media}
As detailed in \textit{Materials and Methods}, we prepare model porous media by confining $\sim10~\mu$m-diameter hydrogel particles, swollen in liquid bacterial culture, at prescribed jammed packing fractions in transparent chambers. Each chamber is mounted on a temperature-controlled stage that maintains a constant temperature throughout to within $\pm1^\circ$C. The individual particle mesh size is $\sim100~$nm, large enough to allow nutrient and oxygen transport throughout but small enough to act as a solid surface to the bacteria.\cite{tapa18,tapa16,tapa18b,tapa15} The hydrogel particle packings therefore act as solid matrices with macroscopic interparticle pores that the individual bacteria can swim through, as schematized in Figure 1a. 

Notably, because the hydrogel particles are highly swollen, light scattering from their surfaces is minimal. As a result, these porous media are transparent, enabling direct visualization and tracking of fluorescent bacteria \textit{in situ} via confocal microscopy. We accomplish this by dispersing a dilute suspension of \textit{E. coli} in the pore space and tracking the center $\vec{\textbf{r}}(t)$ of each cell, projected in two dimensions (2D), over time $t$ with a time resolution of $\delta t=50$, $69$, $100$, or $200$ ms as described further in \textit{Materials and Methods}. 

\begin{figure}[h]
\centering
  \includegraphics[height=3.8cm]{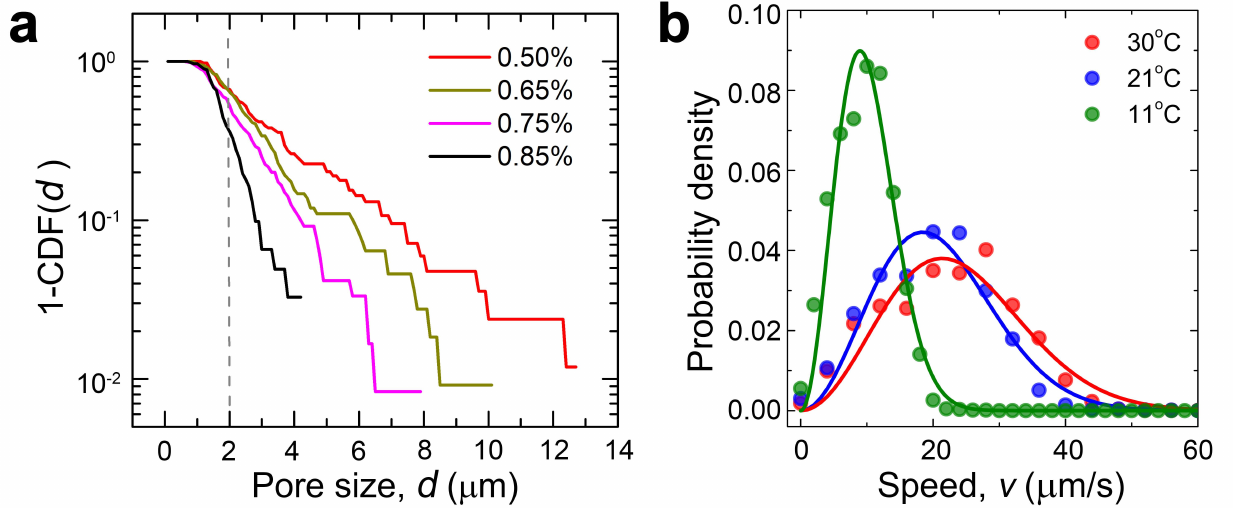}
  \caption{\textbf{Varying confinement and activity by varying pore size and cellular swimming speeds, respectively. a} Complementary cumulative distribution function 1-CDF of the smallest measured pore dimension $d$ measured using thermally-diffusing fluorescent tracers for four different porous media with four different hydrogel particle packing densities. Percentage indicates the dry mass fraction of hydrogel grains used to prepare each medium. Pore sizes are exponentially distributed, as indicated by straight lines on semi-logarithmic scales. The typical cell body length of \textit{E. coli} is indicated by the dashed line. \textbf{b} Probability density of instantaneous cell swimming speeds $v$ measured in homogeneous liquid media at different ambient temperatures. Solid lines indicate Maxwell-Boltzmann distributions fit to the data.}
  \label{fgr:example}
\end{figure}

\begin{figure*}[h]
\centering
  \includegraphics[height=8.3cm]{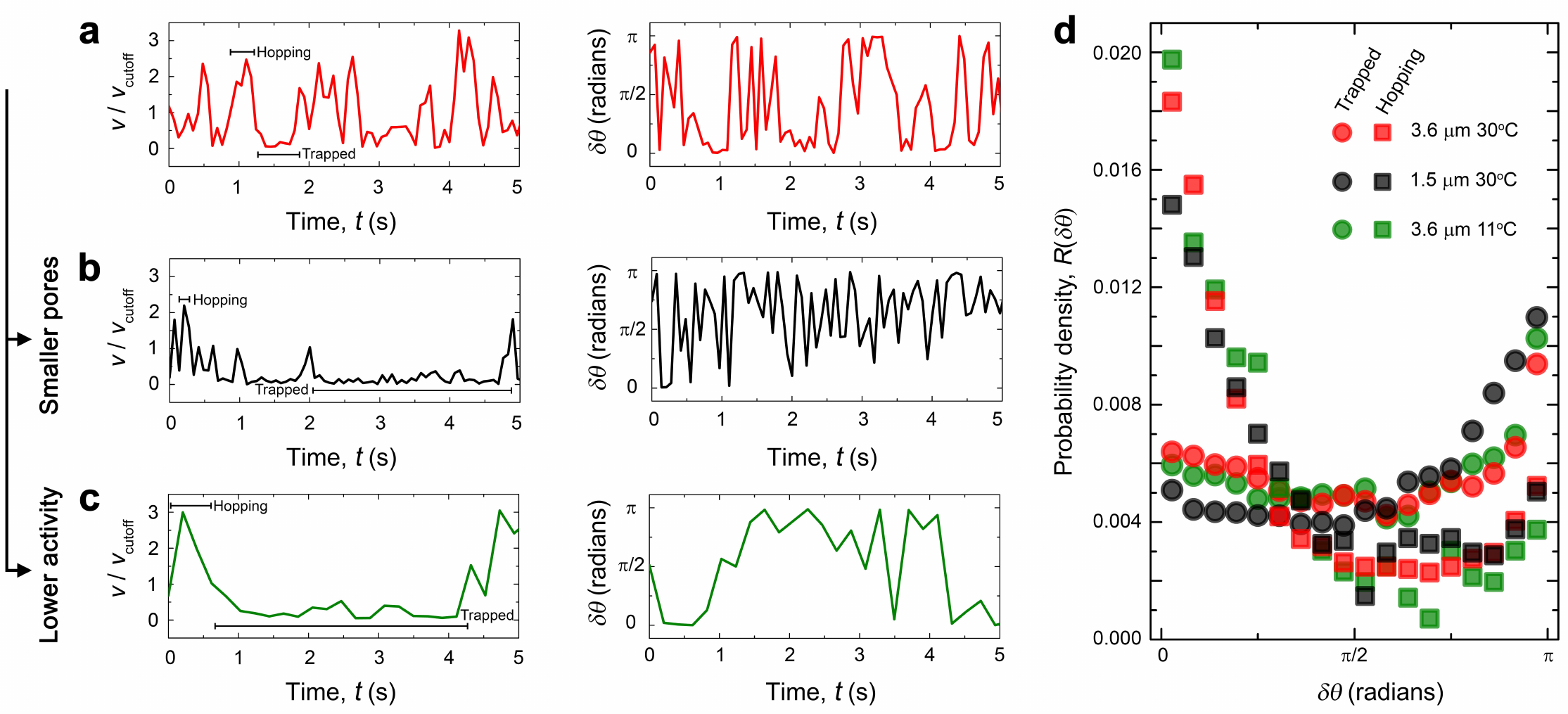}
  \caption{\textbf{Differentiating hopping and trapping by analyzing individual cell trajectories.} First and second columns show temporal traces of instantaneous cell speeds $v$ and reorientation angles $\delta\theta$, respectively, for \textbf{a} large-pore media at 30$^\circ$C, \textbf{b} media with smaller pores at 30$^\circ$C, and \textbf{c} large-pore media at 11$^\circ$C. Speeds are normalized by a cutoff value $v_\text{cutoff}$ equal to half the most probable run speed measured in homogeneous liquid media. Consistent with the trajectories shown in Fig. 1, hopping is shorter and trapping is longer in \textbf{b}, while only trapping is longer in \textbf{c}, compared to \textbf{a}. \textbf{d} In all cases, we find that the probability density of reorientation angles $R(\delta\theta)$ is peaked near $\delta\theta=0$ for hopping, indicating directed motion, while $R(\delta\theta)$ is distributed over all values of $\delta\theta$ for trapping, indicating undirected motion of the cell center. }
  \label{fgr:example}
\end{figure*}

We first investigate the motion of bacteria at 30$^\circ$C, in a medium with pores ranging from $1$ to $13$ $\mu$m in their smallest dimension $d$, as shown by the red line in Fig. 2a; the characteristic pore size is $3.6~\mu$m. Details of the pore size measurement procedure are in \textit{Materials and Methods}. Similar pore size ranges arise in many natural bacterial habitats.\cite{dullien,fatin,lang,zalc,lindquist} Importantly, these pore sizes are comparable to the cell body length $\sim2~\mu$m as indicated by the dashed line in Fig. 2a. For many pores, these sizes are also smaller than the total flagellum length $\sim7~\mu$m; thus, cells can become transiently trapped in tight or tortuous spots as they navigate the pore space. Indeed, as we previously reported,\cite{tapa19} we observe two distinct motility modes that the cells intermittently switch between: hopping, in which a cell moves through a directed path in the pore space, and trapping, in which the cell is confined to a tight spot for extended periods of time. A representative trajectory is shown in Fig. 1b. This motility behavior is in stark contrast to the paradigm of run-and-tumble motion observed in homogeneous media.

To differentiate between hopping and trapping, we analyze the instantaneous speed $v(t)=|\vec{\textbf{v}}(t)|\equiv\left|\vec{\textbf{r}}(t+\delta t)-\vec{\textbf{r}}(t)\right|/\delta t$ of the individual cells as they move. Consistent with the intermittency observed in the cell trajectory in Fig. 1b, we observe intermittent switching between intervals of fast motion and intervals of little to no motion---hopping and trapping, respectively. An example is shown in the first column of Fig. 3a, with the instantaneous speed normalized by a cutoff value $v_\text{cutoff}$, which we set to be $14~\mu$m/s, half the most probable run speed measured in homogeneous liquid media at this temperature; the probability density of measured run speeds in homogeneous media is shown by the red points in Fig. 2b. Following our previous work,\cite{tapa19} we define hopping as a time interval during which $v>v_\text{cutoff}$, corresponding to a minimum displacement $\left|\vec{\textbf{r}}(t+\delta t)-\vec{\textbf{r}}(t)\right|$ of at least $1~\mu$m, the smallest measured pore size, in these experiments. Conversely, trapping is an interval during which $v<v_\text{cutoff}$.

A cell maintains its direction of motion as it hops through the pore space; by contrast, when it is trapped, the cell constantly reorients itself until it can hop again. To quantify this difference in the directedness of motion, we calculate the velocity reorientation angle $\delta\theta(t)\equiv\text{tan}^{-1}\left[\vec{\textbf{v}}(t)\times\vec{\textbf{v}}(t+\delta t)/\vec{\textbf{v}}(t)\cdot\vec{\textbf{v}}(t+\delta t)\right]$ for each cell; a small value of $\delta\theta(t)$ indicates directed motion while larger values reflect increasing amounts of reorientation. As expected, $\delta\theta(t)$ also exhibits intermittent switching between intervals of small $\delta\theta\approx0$ and intervals of large $\delta\theta\approx\pi$ rads, corresponding to hopping and trapping, respectively; a representative example is shown in the second column of Fig. 3a. We further quantify this correspondence using the probability density function of reorientation angles measured for all the tracked cells, $R(\delta\theta)$, calculated separately for hopping and trapping. During hopping, $R(\delta\theta)$ is peaked at $\delta\theta\approx0$, as shown by the red squares in Fig. 3d, indicating directed motion. By contrast, during trapping, $R(\delta\theta)$ is distributed over all values of $\delta\theta$, as shown by the red circles in Fig. 3d, indicating that the cell body randomly reorients itself.

For a cell to hop, it must be able to move along a directed path in the pore space; thus, we hypothesize that the hopping lengths $L_\text{h}$ are determined by the geometry of the medium. We quantify this expectation using the chord length distribution function $\Xi(L_\text{h})$, a fundamental structural metric that describes the probability of finding a straight chord of length $L_\text{h}$ that fits within the pore space;\cite{torquato,lu92} thus, we expect that the distribution of hopping lengths is given by $\Xi(L_\text{h})$. To test this idea, we use direct visualization of the pore space structure to measure $\Xi(L_\text{h})$ as described in \textit{Materials and Methods}, and use our measurements of single-cell trajectories to determine the probability density of hopping lengths $Q(L_\text{h})$. We find good agreement between $\Xi(L_\text{h})$ and $Q(L_\text{h})$, shown by the red line and red points in Fig. 4a, respectively. This agreement suggests that hopping is indeed determined by the geometry of the porous medium. 

\begin{figure*}[h]
\centering
  \includegraphics[height=5.9cm]{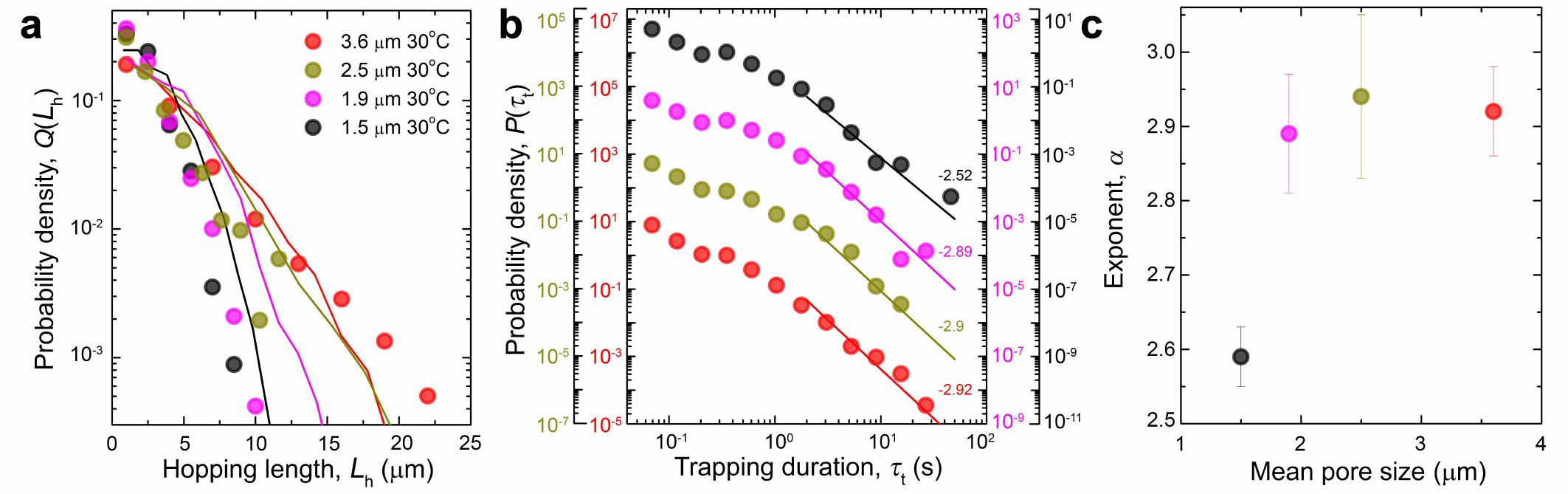}
  \caption{\textbf{Properties of hopping and trapping for varying pore-scale confinement. a} Probability density of measured hopping lengths $Q(L_\text{h})$ in porous media of different pore sizes (circles), all maintained at $30^\circ$C. The legend indicates the characteristic pore size. Solid lines show the measured chord length distribution $\Xi(L_\text{h})$ for each medium. Both $Q(L_\text{h})$ and $\Xi(L_\text{h})$ narrow as the characteristic pore size is decreased, and we find reasonable agreement between the two. \textbf{b} Probability density of measured trapping durations $P(\tau_\text{t})$ in porous media of different pore sizes (circles). The different measurements in different media are offset vertically for clarity; each data set has its own color-coded axis. Lines indicate power-law scaling for large $\tau_\text{t}$, $P(\tau_\text{t})\sim\tau_\text{t}^{-\alpha}$. \textbf{c} The measured power-law exponent $\alpha$ decreases with decreasing characteristic pore size, indicating a broader distribution of trapping durations.}
  \label{fgr:example}
\end{figure*}

We next focus on the dynamics of bacterial trapping in tight spots of the pore space. Our previous single-cell imaging,\cite{tapa19} as well as the data in Fig. 3a-b, indicate that during trapping, a cell constantly reorients itself until it can escape and continue to hop through the pore space. Hence, single cells can be trapped for long durations of time $\tau_\text{t}$, up to $\approx10$ times longer than the run duration in homogeneous media. Our measurement of the full probability density of trapping durations $P(\tau_\text{t})$ is shown by the red points in Fig. 4b. As expected, the trapping durations are distributed over a broad range of $\tau_\text{t}$ spanning nearly three decades. Moreover, $P(\tau_\text{t})$ shows intriguing power-law scaling for large $\tau_\text{t}$: $P(\tau_\text{t})\sim\tau_\text{t}^{-\alpha}$ as indicated by the red line in Fig. 4b. To determine the power-law exponent $\alpha$, we calculate the complementary cumulative distribution function of trapping durations $1-\text{CDF}(\tau_\text{t})\equiv1-\sum_{\tau'_\text{t}<\tau_\text{t}}P(\tau'_\text{t})$ and fit the last two decades in $1-\text{CDF}(\tau_\text{t})$ with a power-law decay. This protocol provides a way to directly determine $\alpha$ free from binning artifacts that arise when directly fitting $P(\tau_\text{t})$, whose shape can change when different bin widths are used. Specifically, for a continuous $P(\tau_\text{t})\sim\tau_\text{t}^{-\alpha}$ with $\alpha>1$, the complementary cumulative distribution function $1-\text{CDF}(\tau_\text{t})=\int_{\tau_\text{t}}^{\infty}P(\tau'_\text{t})\text{d}\tau'_\text{t}\sim\tau_\text{t}^{-(\alpha-1)}$; therefore, fitting the decay of $1-\text{CDF}(\tau_\text{t})$ directly yields $\alpha-1$ without any subjective choice of bin width. We find $\alpha=2.92\pm0.06$, as shown next to the red line in Fig. 4b.

\subsection{Entropic trapping model for bacteria in porous media}

The power-law decay in $P(\tau_\text{t})$ is strikingly similar to that found for trapping of passive species during transport in diverse other disordered systems:\cite{bouchaud} prominent examples include charges in amorphous electronic materials, colloidal particles in dense suspensions or polymer networks, adsorbing solutes in porous media, macromolecules inside cells, and molecules at membranes.\cite{scher,weeks02,wong,drazer,akimoto,yamamoto13,yamamoto14} In all of these cases, the species being transported navigates a disordered landscape of traps of varying confining depths; within each trap, the species is confined until thermal energy $k_{B}T$ enables it to escape. Hence, in these cases, $P(\tau_\text{t})$ decays as $\sim\tau_\text{t}^{-(1+T/T_\text{g})}$ for large $\tau_\text{t}$, where $k_\text{B}$ is Boltzmann's constant, $T$ is ambient temperature, and the mean trapping duration diverges at the temperature $T_\text{g}$---thus, at this point, a dynamical phase transition analogous to a glass transition is thought to occur.\cite{bouchaud} Specifically, the mean trapping duration $\sim\int_{\tau_\text{min}}^{\infty}\tau'_\text{t}P(\tau'_\text{t})\text{d}\tau'_\text{t}\sim\left[\tau_\text{t}^{1-T/T_\text{g}}\right]_{\tau_\text{min}}^{\infty}$, which diverges for $T\leq T_\text{g}$, as described further in the \textit{Appendix}; here $\tau_\text{min}$ is the minimum value of $\tau_\text{t}$ for which $P(\tau_\text{t})$ shows power-law scaling. Power-law scaling of $P(\tau_\text{t})$ for large $\tau_\text{t}$ is therefore thought to be a hallmark of activated transport in a disordered environment, with the power-law exponent signifying how close the system is to a glass-like state.\cite{bouchaud} Motivated by our observations of intermittent trapping with power-law scaling in the tail of $P(\tau_\text{t})$, we hypothesize that bacterial motion in porous media can be understood within a similar theoretical framework---despite the fact that bacteria consume energy, and are thus out of thermal equilibrium. 

First, we consider the dynamics of a trapped cell. As we showed previously,\cite{tapa19} and as supported by the data in Fig. 3a-b, a trapped cell constantly reorients itself until it can find an orientation that enables it to escape and continue to hop through the pore space. This reorientation process is similar to the process by which large polymers thermally escape from tight pores in a disordered porous medium:\cite{muthu87,muthu89a,muthu89b} within such a pore, a polymer chain continually changes its configuration, driven by thermal energy, until one of its ends can escape.  Due to the confining structure of the pore and the molecular properties of the polymer, there are $\Omega_\text{t}$ chain configurations, or trapped states, that keep the polymer trapped within the pore. By contrast, there are only $\Omega_\text{e}$ chain configurations, or transition states, that enable one of the ends to escape; this transition state then enables the polymer to continue to diffuse through the pore space. The entropies of the trapped and transition states are then given by $k_\text{B}\text{ln}~\Omega_\text{t}$ and $k_\text{B}\text{ln}~\Omega_\text{e}$, respectively, and the free energy difference between them is given by $C\equiv k_\text{B}T~\text{ln}\left(\Omega_\text{t}/\Omega_\text{e}\right)$.\cite{han} The pore can therefore be thought of as an "entropic trap" with trap depth $C$. Because the process by which the polymer escapes the trap is thermally-activated, the probability for the polymer to escape is given by an Arrhenius-like relation; the trapping duration is then $\tau_\text{t}\sim e^{C/k_{B}T}$. By analogy, for the case of bacteria moving through the pore space, we assume that tight spots of the pore space can again be described by trap depths $C$---which likely depend on the pore size and structure, the cell morphology and elastic properties, and any cell-solid surface interactions---with $\Omega_\text{t}$ and $\Omega_\text{e}$ now representing the orientations of the cell that keep it trapped within or enable it to escape from each pore, respectively. Moreover, because bacteria are active, we replace the thermal energy $k_\text{B}T$ by the cellular activity $X$, which describes how actively the cell reorients and attempts to escape from the trap.\cite{Woillez2019,Geiseler2016,Bi2016,Takatori2015} The trapping duration is then given by $\tau_\text{t}=\tau_0 e^{C/X}$, where $\tau_{0}>0$ is the minimum trapping duration due to pore-scale confinement alone.

The pore space is disordered, with a broad, exponential distribution of pore sizes, as shown in Fig. 2a; we thus assume that the trap depths are exponentially distributed, as is found in diverse other disordered media,\cite{bouchaud} with probability density $\rho(C)=C_{0}^{-1}e^{-C/C_{0}}$. The parameter $C_{0}$ characterizes the mean trap depth within a given medium. The probability density of trapping durations is then given by $P(\tau_\text{t})=\rho(C)/\left(\partial\tau_\text{t}/\partial C\right)=\left(\beta/\tau_\text{t}\right)e^{-C/C_0}=\beta\tau_{0}^{\beta}\tau_\text{t}^{-(1+\beta)}$ for $\tau_\text{t}>\tau_{0}$; in this case the parameter $\beta\equiv X/C_0$, analogous to $T/T_\text{g}$ for thermally-equilibrated systems, characterizes the competition between cellular activity and pore-scale confinement in the medium. Hence, the probability density of trapping durations $P(\tau_\text{t})\sim\tau_\text{t}^{-\alpha}$ for sufficiently large $\tau_\text{t}$ as in the experiments, with $\alpha\equiv1+\beta$. Within this theoretical framework, we therefore expect that $P(\tau_\text{t})$ shows power-law scaling for diverse disordered porous media, with an exponent whose magnitude decreases with increasing pore-scale confinement $C_0$ or decreasing cellular activity $X$---approaching a glass-like state with a diverging mean trapping duration when $X\leq C_0$. The mathematical details of this divergence are given in the \textit{Appendix}. To test this hypothesis, we use media with different pore sizes to first vary $C_0$, and use bacteria with different swimming speeds to independently vary $X$, as described in Sections 2.3 and 2.4, respectively. 

\begin{figure}[h]
\centering
  \includegraphics[height=3.4cm]{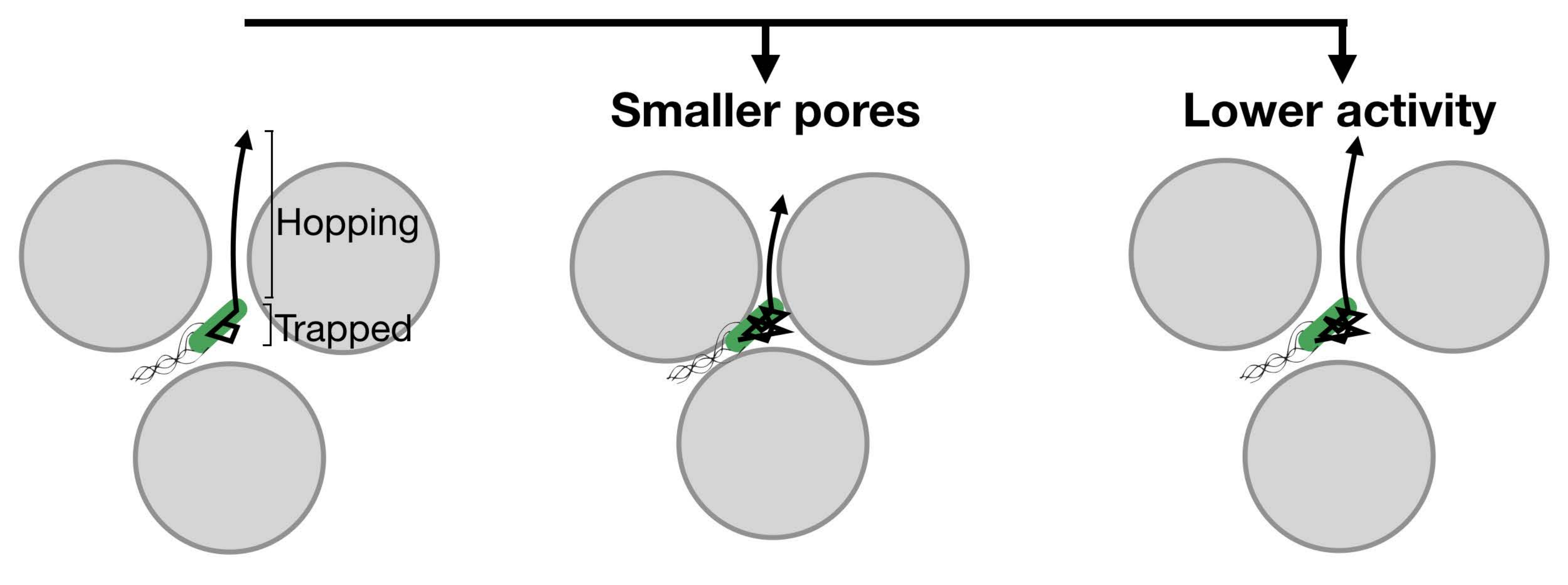}
  \caption{\textbf{Schematic illustrating the separate influences of pore-scale confinement and cellular activity on bacterial motion in porous media.} Solid grains are shown by grey circles. As the pores are made smaller (second column), the cell is increasingly confined, resulting in shorter hopping lengths and longer trapping durations. As the cellular activity is reduced (third column), the hopping length is unchanged, but trapping durations are longer.}
  \label{fgr:example}
\end{figure}

\subsection{Influence of pore-scale confinement}
We tune the pore size distribution of the media, and therefore the degree of confinement $C_0$, by varying the hydrogel particle packing density: increasing the packing density yields media with characteristic pore sizes of $2.5$, $1.9$, and $1.5$ $\mu$m, indicated by the yellow, magenta, and black lines in Fig. 2a, respectively. As schematized in Fig. 5, we expect that in media with smaller pores, the hopping lengths $L_\text{h}$---which we hypothesize are determined by the geometry of the medium---will be smaller, and the trapping durations $\tau_\text{t}$ will be larger. Consistent with this expectation, we find that the bacteria still move via intermittent hopping and trapping, but with shorter hops and longer trapping durations; a representative trajectory is shown in Fig. 1c. This behavior is quantified in Fig. 3b and by the black points in Fig. 3d: we again observe intermittent switching between intervals of fast, directed motion and slow, undirected motion---hopping and trapping, respectively. Compared to media with larger pores, however, the intervals of hopping are shorter, and the intervals of trapping are longer: compare Fig. 3b to 3a.  

To further quantify this behavior, we first investigate the influence of pore-scale confinement on bacterial hopping lengths. We again directly measure the chord length distribution function $\Xi(L_\text{h})$, as well as the probability density of bacterial hopping lengths $Q(L_\text{h})$, for each medium. As expected, both $\Xi(L_\text{h})$ and $Q(L_\text{h})$ narrow, extending to smaller maximal values of $L_\text{h}$, as pore size is decreased. Moreover, we find reasonable agreement between the measured $\Xi(L_\text{h})$ and $Q(L_\text{h})$, shown by the lines and points in Fig. 4a, respectively, further supporting the idea that hopping is determined by the geometry of the porous medium.

We next investigate the influence of pore-scale confinement on bacterial trapping durations. As described in Section 2.2, increasing the degree of confinement $C_0$ should result in a broader distribution of $\tau_\text{t}$, characterized by a smaller value of the exponent $\alpha$. We test this expectation by again measuring the probability density of trapping durations $P(\tau_\text{t})$ for individual bacteria swimming in each medium. Consistent with our entropic trapping model, we observe broad distributions of $\tau_\text{t}$, with extended tails consistent with power-law scaling for large $\tau_\text{t}>\tau_{0}\approx2$ s: $P(\tau_\text{t})\sim\tau_\text{t}^{-\alpha}$ as shown by the lines in Fig. 4b. Moreover, the power-law exponent determined from the complementary cumulative distribution function decreases with decreasing pore size, consistent with the prediction that $\alpha=1+X/C_0$; this behavior is summarized in Fig. 4c. Intriguingly, $\alpha$ appears to decrease more precipitously as pore size decreases below the cell body length $\approx2~\mu$m, indicating that confinement plays a more dominant role in this regime. Together, these results thus support the idea that trapping is regulated by pore-scale confinement, and can be described by an entropic trapping model.

\begin{figure*}[h]
\centering
  \includegraphics[height=5.9cm]{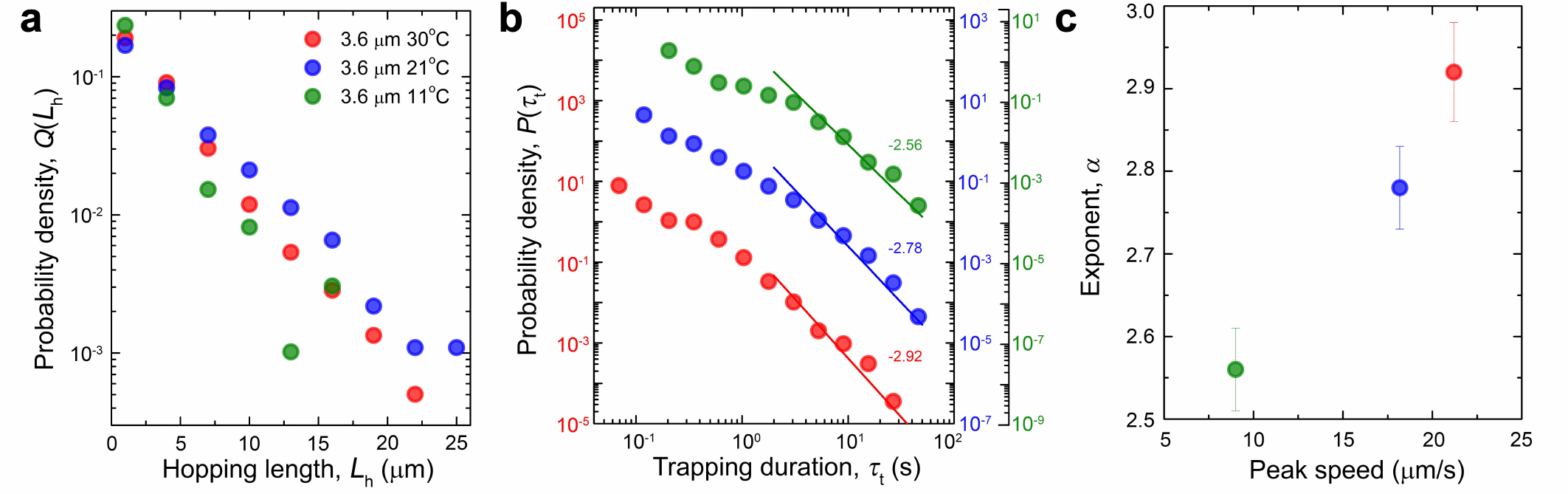}
  \caption{\textbf{Properties of hopping and trapping for varying cellular activity. a} Probability density of measured hopping lengths $Q(L_\text{h})$ in porous media maintained at different temperatures, all with the same characteristic pore size. The legend indicates the ambient temperature. We do not find any systematic variation of $Q(L_\text{h})$ with temperature. \textbf{b} Probability density of measured trapping durations $P(\tau_\text{t})$ in porous media maintained at different temperatures (circles). The different measurements at different temperatures are offset vertically for clarity; each data set has its own color-coded axis. Lines indicate power-law scaling for large $\tau_\text{t}$, $P(\tau_\text{t})\sim\tau_\text{t}^{-\alpha}$. \textbf{c} The measured power-law exponent $\alpha$ decreases with decreasing temperature, and thus with mean swimming speed, indicating a broader distribution of trapping durations. Horizontal axis shows the peak speed obtained from Maxwell-Boltzmann fits in Fig. 2 at each temperature tested.}
  \label{fgr:example}
\end{figure*}

\subsection{Influence of cellular activity}
Another key parameter that may influence bacterial motion in porous media is how actively the cells swim through the pore space. We tune the bacterial swimming speed, and therefore the cellular activity $X$, by changing the ambient temperature. Previous studies have shown that decreasing temperature below $30^\circ$C decreases the bacterial run speed, and increases the run duration, in homogeneous media.\cite{Maeda1976} Consistent with these results, we find that the run speeds in homogeneous media decrease as temperature is decreased to as low as $11^\circ$C, as shown by the blue and green points in Fig. 2b; the run durations concomitantly increase, leading to run lengths that do not vary considerably with temperature, as described further in \textit{Materials and Methods}. Ambient temperature thus provides a straightforward way to tune cellular activity. 

To isolate the influence of cellular activity on bacterial motion, we use three different porous media having the same pore size distribution, all with a characteristic pore size of $3.6~\mu$m, at three different temperatures. As schematized in Fig. 5, we expect that for bacteria with lowered activity, the hopping lengths $L_\text{h}$---which we hypothesize are determined solely by the geometry of the medium---will be unchanged, while the trapping durations $\tau_\text{t}$ will be larger. Consistent with this expectation, we find that cells again move via intermittent hopping and trapping, with hops of similar lengths but punctuated by longer intervals of trapping; a representative trajectory is shown in Fig. 1d. This behavior is quantified in Fig. 3c and by the green points in Fig. 3d: we again observe intermittent switching between intervals of fast, directed motion and intervals of slow, undirected motion---hopping and trapping, respectively. Compared to bacteria with higher activities, however, the intervals of trapping are longer: compare Fig. 3c to 3a.  

To further quantify this behavior, we first investigate the influence of cellular activity on bacterial hopping lengths. We again directly measure the probability density of bacterial hopping lengths $Q(L_\text{h})$ for each temperature tested. The measurements show slight variations due to variability in preparation of the media. However, in stark contrast to the case of increasing confinement, we do not observe a systematic variation of $Q(L_\text{h})$ with temperature. Instead, as expected, $Q(L_\text{h})$ appears to be temperature-independent, as shown by the different colors in Fig. 6a. These data further confirm the idea that hopping is determined solely by the geometry of the porous medium; hopping lengths do not depend on cellular activity.

We next investigate the influence of cellular activity on bacterial trapping durations. As described in Section 2.2, decreasing the activity $X$ should result in a broader distribution of $\tau_\text{t}$, characterized by a smaller value of the exponent $\alpha$. We test this expectation by again measuring the probability density of trapping durations $P(\tau_\text{t})$ for individual bacteria swimming in each medium. Consistent with our entropic trapping model, we observe broad distributions of $\tau_\text{t}$, with extended tails consistent with power-law scaling for large $\tau_\text{t}>\tau_{0}\approx2$ s in all cases: $P(\tau_\text{t})\sim\tau_\text{t}^{-\alpha}$ as shown by the lines in Fig. 6b. Moreover, the power-law exponent determined from the complementary cumulative distribution function decreases with decreasing cellular activity, consistent with the prediction that $\alpha=1+X/C_0$; this behavior is summarized in Fig. 6c. These results thus support the idea that trapping is also regulated by cellular activity, and can be described by our entropic trapping model.

\section{Conclusions}
Our experiments characterize how bacterial motion in porous media is regulated by both pore-scale confinement and cellular activity. By preparing media with varying pore sizes, we isolate the influence of confinement, while by tuning the environmental temperature and thus the bacterial swimming speed, we independently isolate the influence of cellular activity. Our work explores the case of tight porous media, characteristic of many bacterial habitats, having pore sizes comparable to or smaller than the overall cell size---and much smaller than the homogeneous run length. Thus, for all conditions tested, we find that bacteria exhibit hopping-and-trapping motility, with faster, directed hops punctuated by slower, prolonged, undirected intervals of trapping. However, the hopping lengths and trapping times depend sensitively on porous medium geometry and environmental conditions.

Our results suggest that hopping is determined solely by the geometry of the medium---specifically, by the availability of straight paths in the pore space---and is not modulated by variations in cellular activity. By contrast, we find that trapping is determined by the competition between pore-scale confinement, which promotes trapping, and cellular activity, which suppresses trapping. This work thus expands on our previous discovery of hopping-and-trapping motility in porous media, shedding light on the factors that control this new form of motility. We anticipate that our data could help to test current models, and could motivate the development of new models, of motility in heterogeneous environments.\cite{roseanne95,croze,hilpert,licata,lushi,morin,vasily,reich1,Bechinger2016b,Mokhtari2019}
  
Indeed, the process of hopping and trapping bears striking similarities with the entropic trapping of thermally-activated polymers in disordered porous media; thus, by analogy, we hypothesize that bacterial trapping can be described by the parameter $\beta\equiv X/C_0$, analogous to $T/T_\text{g}$ for thermally-equilibrated systems. Consistent with this hypothesis, we find that the power-law exponent $\alpha\equiv1+\beta$ decreases with decreasing pore size, which increases $C_0$, and with decreasing swimming speed, which decreases $X$. We thus expect that $\alpha$ will further decrease with increasing $C_0$ or decreasing $X$, eventually reaching 2 when $X=C_0$; at this point, the mean trapping duration diverges, and we hypothesize that bacterial motion approaches a glass-like state. Testing this prediction will be an important direction for future research. Moreover, how exactly to determine $C_0$ and $X$---which likely depend on a complex interplay between pore structure, cell-surface interactions, cellular morphology/mechanics and surface properties, and cellular swimming kinematics\cite{lushi,molaei,ford09,ford11,Woillez2019,Geiseler2016,Bi2016,Takatori2015}---remains an outstanding theoretical question.

\section{Materials and Methods}
\subsection{Preparation of transparent 3D porous media}
To prepare a 3D porous medium, we disperse a fixed mass fraction of dried, cross-linked, biocompatible acrylic acid/alkyl acrylate copolymer grains (Carbomer 980, Ashland) in liquid LB media (Lennox Lysogeny Broth). We ensure a homogeneous dispersion by mixing for at least 2 h. The grains then swell to form a jammed, disordered packing comprised of $\sim10~\mu$m-diameter hydrogel particles, with $\approx20\%$ polydispersity, as determined previously by others using phase contrast microscopy, light scattering, and image correlation spectroscopy of hydrogel particles similar to those we use in aqueous solvents.\cite{tapa18b,epje,lee} We estimate the effective packing density as $\phi_\text{eff}=\phi_\text{j}(C/C_\text{j})$, where $C$ is the total polymer mass fraction and $\phi_\text{j}\approx0.64$ is the packing volume fraction at the onset of jamming with a corresponding polymer mass fraction that we determine to be approximately $C_\text{j}\approx0.5\%$ via shear rheology. Thus, for the four different porous media used here with $C=$ 0.5, 0.65, 0.75, and 0.85\%, the effective packing fractions are $\phi_\text{eff}\approx$ 0.64, 0.83, 0.96 and 1.09, respectively. We note, however, that we measure connected void space between hydrogel particles even in the densest packings with $\phi_\text{eff}\approx1$, which presumably reflects the influence of polydispersity and possible deswelling of the individual hydrogel particles at the highest packing fractions.

We confine 4 mL of each packing in a transparent, sealed glass-bottom petri dish (Cellvis) $\approx3.5$ cm in diameter and $\approx1$ cm thick with an overlying thin layer of $750~\mu$L LB media or $1$ mL paraffin oil to minimize evaporation. We adjust the final pH to 7.4 by adding $10$ N NaOH. The internal mesh size $\xi$ of each particle is $\sim100$ nm, as inferred from shear rheology measurements of the elastic modulus $G'$ of the jammed packings: specifically, $\xi\approx(k_{B} T/G')^{1/3}$, with $k_B$ representing Boltzmann's constant,  $T$ representing temperature, and we measure $G'\approx5$ Pa near the onset of jamming.\cite{tapa18b,ewoldt,polymer} This mesh size is much smaller than the individual bacteria, but large enough to allow unimpeded transport of nutrients and oxygen.\cite{tapa18,tapa16,tapa15} Moreover, because the bulk modulus of the individual particles\cite{Lietor-Santos2011} $\sim10^3$ Pa is larger than the stress generated by swimming, $\sim~10^{-2}$ Pa at Reynolds numbers $\sim10^{-5}$, the packings act as solid matrices with macroscopic interparticle pores that the cells can swim through. Because the individual cells move through the void space between the packed hydrogel particles, we expect that they do not compress the hydrogel particles, nor are they compressed by the hydrogel particles. We expect similar behavior for the tracer particles used, which are 200 nm in diameter---an order of magnitude smaller than the \textit{E. coli} cell body length and a factor of 5 smaller than the smallest pore size we measure. 

Notably, because the hydrogel particles are highly swollen, light scattering from their surfaces is minimal. As a result, these porous media are transparent, enabling direct visualization and tracking of fluorescent particles and bacteria \textit{in situ} via confocal microscopy. This approach thus overcomes the limitation of typical opaque media. Specifically, we use a Nikon A1R inverted laser-scanning confocal microscope with a temperature-controlled stage for all experiments described. To avoid boundary effects, all imaging is performed at least $100~\mu$m from the bottom of the chamber containing the medium.

We note that the osmotic pressure of the liquid LB media is $\sim1$ MPa, due to the high concentration of dissolved salts. This value is $\sim10^{3}$ times larger than the hydrogel bulk modulus, more than sufficient to compress the particles; yet, the hydrogel particles still swell in the LB media. We speculate that the reason why the hydrogel particles can still swell in the LB media is due to their large mesh size, which enables free diffusion of salt within each hydrogel particle and minimizes the difference in salt concentration between the hydrogel interior and exterior. As a result, the osmotic pressure difference between the hydrogel interior and exterior is minimal, and each particle can still swell, despite the salinity of the LB media used. 

\subsection{Characterizing the pore space geometry}
Tuning the hydrogel particle packing density provides a way to tune the pore size distribution. To measure the pore size distribution of each medium, we disperse a dilute suspension of $200$ nm diameter fluorescent tracers (carboxylated polystyrene FluoSpheres, Invitrogen) in the pore space and track their thermal diffusion. The tracer zeta potential is approximately $-20$ mV, comparable to that of \textit{E. coli}, approximately $-30$ mV. Since the tracers are larger than the hydrogel mesh size, they only diffuse through the interparticle pore space. We track the center of each tracer at a temporal resolution of $50$ ms using a custom MATLAB script: we identify each center using a peak finding function with subpixel precision and track its motion via the classic Crocker-Grier algorithm.\cite{crocker} This tracking enables us to determine the mean-squared displacement (MSD) of each tracer as it explores the pore space as a function of lag time.  At short lag times, a tracer diffuses unimpeded within the pore space; hence, the MSD varies linearly in time. Over sufficiently long lag times, however, the tracer motion is impeded by interactions with the pore walls, and the MSD plateaus. We identify this plateau value for each measured MSD, and calculate the smallest confining pore dimension $d$ as the sum of the square root of this plateau value and the tracer diameter. To describe the distribution of the pore sizes $d$ for each medium, we calculate the complementary cumulative distribution function $1-\text{CDF}(d)\equiv1-\sum_{d'<d}p(d')$, where $p(d)$ is the probability density of $d$. These measurements are shown in Fig. 2a and indicate that the pore sizes are exponentially distributed, shown by the linear scaling of $1-\text{CDF}(d)$ on semi-logarithmic axes.

To measure the chord length distribution function $\Xi(L_\text{h})$ for each medium, we construct maximum-intensity projections of tracer motion, for each tracer, over the entire imaging duration. These projections provide a map of the pore space. We then binarize the projections and directly calculate the distribution of chords of length $L_\text{h}$ that fit within each projected pore space image. This protocol yields a direct measurement of $\Xi(L_\text{h})$.

\subsection{Imaging bacterial motion in the pore space}
In this study, we use \textit{E. coli} (strain W3110), a model microswimmer that exhibits run-and-tumble motion in homogeneous liquid media. The cells constitutively express green fluorescent protein (GFP) throughout their cytoplasm, facilitating fluorescent visualization via confocal microscopy. In each experiment, we prepare a $1:100$ dilution of an overnight culture in fresh LB media, and culture the resulting mixture of cells at $30^\circ$C for $3$ h until the optical density is $\approx0.6$, corresponding to exponential growth. A small volume, $20~\mu$L, of the resulting culture is then homogeneously dispersed within the pore space of a given medium by gentle pipette mixing to a total bacterial concentration of $8000$ cells/$\mu$L, sufficiently dilute to minimize intercellular interactions, changes in oxygen content, or changes or nutrient content within the medium. The resulting dilution in the quoted hydrogel mass fraction is negligible---less than $0.004\%$. The pipette mixing fluidizes the hydrogel porous medium, enabling the cells to be dispersed within the pore space; the medium then rapidly transitions to a solid state after mixing is completed due to the re-packing of the jammed particles around the cells.\cite{tapa15,tapa16} 

To monitor bacteria motion through the pore space, we use confocal microscopy to acquire projected 2D movies with an optical slice thickness of $79~\mu$m and at a temporal resolution of $50$ or $69$ ms. We estimate that tracking errors arising from projection effects/limitations in image acquisition speed are minimal; only $0.13\%$ of motions are erroneously detected due to projection effects, as we described previously.\cite{tapa19} Similar to the tracking of fluorescent tracers, we use a custom MATLAB script to identify each cell center using a peak finding function with subpixel precision and track its motion via the classic Crocker-Grier algorithm.\cite{crocker} We track the motion of each cell for at least $10$ s and focus our analysis on cells that exhibit motility within the tracked time. The imaging time scale for each individual experiment is $\sim2$ min. This duration is more than double the longest trapping duration we measure, ensuring that our imaging duration is sufficiently long. In addition, this duration is over an order of magnitude shorter than the cell division time under these conditions, $\sim30$ to $40$ min; thus, our measurements of motility are not influenced by cellular growth and division. Importantly, we do not detect any changes in pore structure over the entire experimental time scale, using trapped cells as tracers.

We also perform control experiments in homogeneous liquid LB within a bulk 3D chamber, away from any boundaries. Consistent with our and others' previous measurements,\cite{berg} in homogeneous media, we find that \textit{E. coli} exhibit run-and-tumble motility: they perform ballistic, directed runs punctuated by rapid tumbles that randomly reorient the cells. Thus, the ensemble- and time-averaged MSD varies quadratically in time for short lag times, indicating ballistic runs; at lag times longer than the mean run duration, the MSD varies linearly in time, indicating diffusive behavior. Fitting the short-time MSD therefore yields a measurement of the mean run speed that agrees with the data shown in Fig. 2b. Consistent with previous studies,\cite{Maeda1976} we find that while the run speed decreases with decreasing ambient temperature, the run duration increases: we measure mean run durations of $\approx2$, $2.5$, and $4$ s for $T=30$, $21$, and $11^\circ$C, respectively. Hence, the mean run length $\approx35-45~\mu$m is always much larger than the measured pore sizes in our experiments, and hopping lengths are determined by the geometry of the medium. 

Because the cells move slower at lower $T$, using an excessively small value of the sampling time resolution $\delta t$ exacerbates noise in the calculation of the instantaneous speed $v(t)=|\vec{\textbf{v}}(t)|\equiv\left|\vec{\textbf{r}}(t+\delta t)-\vec{\textbf{r}}(t)\right|/\delta t$ of each cell. To avoid this noise enhancement due to oversampling, we calculate the ensemble- and time-averaged MSD for bacteria in homogeneous liquid media with different choices of $\delta t$; when $\delta t$ is too small, the short-time MSD is dominated by noise and does not vary quadratically with lag time. We identify the optimal value of $\delta t$ to use by identifying the minimal value of $\delta t$ at which the measured MSD varies quadratically with lag time at short times, as expected for ballistic motion; increasing $\delta t$ above this optimal value does not change the measured MSD, indicating that this analysis is robust to subsequent variations in sampling. This analysis protocol thus prevents noise due to oversampling, and yields values of $\delta t=69$, $100$, and $200$ ms for $T=30$, $21$, and $11^\circ$C, respectively. We use these values in all analysis of bacterial motion both in homogeneous liquid media and in porous media at different temperatures, including in our calculations of the instantaneous speed $v(t)$.

\section{Appendix: mean trapping duration}
Here, we calculate the mean trapping duration $\langle\tau_\text{t}\rangle$ for the case in which the probability density of trapping durations is given by $P(\tau_\text{t})=P_{0}\tau_\text{t}^{-\alpha}$ for $\tau_\text{t}$ ranging from $\tau_\text{min}$ to $\infty$; for $0<\tau_\text{t}<\tau_\text{min}$, we simply represent $\int_{0}^{\tau_\text{min}}P(\tau'_\text{t})\text{d}\tau'_\text{t}$ by the non-dimensional constant $A$.

We first note that $P(\tau_\text{t})$ is normalized: 

\begin{equation*}
A+\int_{\tau_\text{min}}^{\infty}P(\tau'_\text{t})\text{d}\tau'_\text{t}=1\implies P_{0}=(1-A)(\alpha-1)\tau_\text{min}^{\alpha-1}
\end{equation*}
with $\alpha>1$.

Then, $\langle\tau_\text{t}\rangle=\tau_{0}+\int_{\tau_\text{min}}^{\infty}\tau'_\text{t}P(\tau'_\text{t})\text{d}\tau'_\text{t}$ where $\tau_{0}\equiv\int_{0}^{\tau_\text{min}}\tau'_\text{t}P(\tau'_\text{t})\text{d}\tau'_\text{t}$ is a finite value. Substituting for $P(\tau'_\text{t})$ yields 

\begin{equation*}
\langle\tau_\text{t}\rangle=\tau_{0}+\frac{P_{0}}{2-\alpha}\left[\tau_\text{t}^{2-\alpha}\right]_{\tau_\text{min}}^{\infty}=\tau_{0}-(1-A)\tau_\text{min}^{\alpha-1}\frac{\alpha-1}{\alpha-2}\left[\tau_\text{t}^{2-\alpha}\right]_{\tau_\text{min}}^{\infty}.
\end{equation*}

\noindent This expression for $\langle\tau_\text{t}\rangle$ diverges when $2-\alpha\geq0$, or when $\alpha\leq2$. Conversely, when $\alpha>2$, the mean trapping duration is well-defined; it is given by

\begin{equation*}
\langle\tau_\text{t}\rangle=\tau_{0}+(1-A)\frac{\alpha-1}{\alpha-2}\tau_\text{min},
\end{equation*}

\noindent which decreases with increasing $\alpha$, eventually converging to $\langle\tau_\text{t}\rangle\approx\tau_{0}+(1-A)\tau_\text{min}
$ as $\alpha\rightarrow\infty$.

\section*{Conflicts of interest}
There are no conflicts to declare.

\section*{Acknowledgements}
This work was supported by start-up funds from Princeton University, the Project X Innovation fund, a distinguished postdoctoral fellowship from the Andlinger Center for Energy and the Environment at Princeton University to T.B., and in part by funding from the Princeton Center for Complex Materials, a Materials Research Science and Engineering Center supported by NSF grant DMR-1420541.



\balance


\providecommand*{\mcitethebibliography}{\thebibliography}
\csname @ifundefined\endcsname{endmcitethebibliography}
{\let\endmcitethebibliography\endthebibliography}{}

\bibliographystyle{rsc} 

\end{document}